# An Efficient Light-weight LSB steganography with Deep learning Steganalysis

Dipnarayan Das, National Institute of Technology, Durgapur, West Bengal, dipnarayan.das35@gmail.com
Asha Durafe, Shah & Anchor Kutchhi, Engineering College, Mumbai, Sir Padampat Singhania University, Udaipur, asha.durafe@sakec.ac.in
Dr. Vinod Patidar, Sir Padampat Singhania University, Udaipur, vinod.patidar@spsu.ac.in

**Abstract:**

Active research is going on to securely transmit a secret message or so-called steganography by using data hiding techniques in digital images. After assessing the state-of-the-art research work, we found, most of the existing solutions are not promising and ineffective against machine learning-based steganalysis. In this paper, a lightweight stegnography scheme is presented through graphical key embedding and obfuscation of data through encryption. By keeping a mindset of industrial applicability, to show effectiveness of proposed scheme, we emphasized mainly on deep learning based steganalysis. The proposed steganography algorithm containing two schemes withstands not only statistical pattern recognizers but also machine learning steganalysis through feature extraction using a well-known pre-trained deep learning network Xception. We provided a detailed protocol of the algorithm at different scenarios and implementation details. Furthermore, different performance metrics are also evaluated with statistical and machine learning performance analysis. The results were quite impressive respect to state of the arts. We received 2.55% accuracy through statistical steganalysis and machine learning steganalysis gave maximum 49.93~50% correctly classified instances at good condition.

## 1. Introduction:

As we discuss the science of secret messaging, one question always comes first, what is the difference between cryptography and steganography. In cryptography [1,2], a method converts secret plain data into cipher data [3] and sents to another person who then decrypts the cipher data into plain data (secret). But in case of steganography, the secret data is unaltered and scrambled inside a media for hiding. Like in image steganography the transmission media aka cover data, is an image and the secret data aka stego data is hidden inside the cover. The main target of steganography is not to protect from decryption but make the transmission unsuspicious. From the definition we got our motivation which is human's curiosity. Today where we are it is mainly effect of curiosity of human which drives us to explore more. From a defender perspective if we can remove the curiosity of hacker then we will win the war before starting. As an example if a hacker tried to login in admin page and get response "incorrect password" then due to curiosity s/he will again try with a new password. But if we give a fake access without stopping, so hacker's focus will not be isolated to only that fake environment. To implement the concept in our steganography algorithm, we proposed obfuscation.

Now we came to know steganography is to conceal the secret data within multimedia contents such as file, message, image, or video. [4-5] whereas is to detect secret data hidden using steganography, identify suspected packages and ascertain whether the secret data is embedded or not [6-8]. An added advantage of steganography over the cryptography is that, in case of cryptography the algorithms and keys have finite range for encryption but in steganography the number of media is infinite and as there are two computations needed (Reading the cover data & Analyzing Secret pattern), the process became more complex.

Though steganography has effective security, today's high-speed computations with the help of AI Automation break most of the state-of-the-art algorithms. Under AI, machine learning is combined with deep learning networks for pattern analysis. The concepts of machine learning [9-10] and artificial intelligence (AI) becoming almost synonymous with information security, more and more enterprises turning to automation and 'cognitive computing to improve the proficiency of their security efforts. However, where there is light there is also dark [11] too. According to recent research from ESET, the threat of AI being used as a weapon against organizations has led to a significant amount of IT decision makers (75%) in the US to believe that the number of attacks they have to detect and respond to will increase.

In 2006, Mielikainen [12] introduced the LSB scheme of embedding two bits into a pixel pair of cover image. Zhang and Wang [13] proposed the EMD data hiding scheme, to overcome the weakness in [12], but the payload only 1.161 bpp for embedding a base-5 numeral system into a pixel pair. In [13], each secret digit in a base-$(2n + 1)$ numerical system was carried by a pixel-group, which achieves a higher embedding capacity and provides better stego-image quality than previous approaches. Kim et al. [14] provided a inventive step to improve the quality of Mielikainen's method [12]. In 2008, Chang et al. [15] presented a Sudoku-based data hiding scheme to improve security, a secret digit in base $-9$ numeral system is embedded into each pair of pixels, but their scheme had a large search set and only could embed base $-9$ numeral system, not flexible enough to multiple embedded requirements. In 2012, Hong and Chen [16] proposed an adaptive pixel pair matching method, where the scrambled plain data can be extracted without the key or additional information. In 2014, J. Chen [17] proposed a pixel pair matching(PPM) with Pixel value difference(PVD) method making the secret data embedded adaptively into pixel pair using two reference tables and had the resistance to chi-square steganalysis. And Chang, Liu and Nguyen [18] used turtle shell matrix. Their scheme increased the embedding rate. It makes flexible embedding version in 2017 [19], and the faster search version and in 2018 [20]. But once the embedding algorithm is revealed any malicious individual could extract the secret information easily, so encrypting the secret data [21] is a necessity for their proposed scheme. Keeping in mind several schemes based on secret keys have been proposed in recent years [22]. Yadav and Ojha [23] proposed a more efficient scheme through chaotic systems for security, which provides a higher payload and imperceptibility. Rasool et. al. [24] mainly focused on the steganalysis of RGB images where a dataset named "CALTECH BMP" is utilized and in the images different steganography algorithms applied to hide the secret message. From cross validation and testing results we can see the exactness of SVM is in between 99% to 100%. The paper proposed that their trained model has given high precision of over 99% for the consolidated RGB channels features with dual channel combinations as well as single channels. The number of different features is not in any dependency. Inserting in one channel in particular (the blue channel) there was no decrease in the recognition accuracy. Joshi et. al. [25] proposed a new method of Image Steganography where the 7th bit of a pixel is working as an indicator and a successive Temporary Pixel in the grayscale image. Rasras et. al. [26] have shown a quite good PSNR value. Qin et. al. [27] proposed a coverless image steganography technique where a region detection system works to find a suitable place to embed secret data into an image to make it undetectable. Durafe and Patidar [28] presented a blind color image steganography with fractal cover and compared the performance metrics with IWT (Integer Wavelet Transform)-SVD (Singular Value Decomposition) and DWT (Discrete Wavelet Transform)-SVD embedding in frequency domain. Qian et. al. [29] proposed a steganography algorithm that can break convolutional neural network based steganalysis tool thus enhancing security of the hidden data.

Reviewing subsequent and more recent literature it was apparent that most of the steganography approach's firewall is broken against machine learning steganalysis. In few scenarios, the mathematical computation can be reversed without complexity. It doesn't mean that the methods of the state of the arts are not good in embedding but according to the extensive analysis done it can be predicted that the embedding is done at specific places of the images which coordinately generate a pattern and which can be exploited by feature engineering. Inspired by state-of-the-art algorithms, to protect embedding from getting an interest, the proposed work utilizes two layers of security, first one is the obfuscation layer which makes the embedding, nearly undetectable and the second layer is graphical embedding which can`t be breakable using visual recognition. As a significance of this study we leveraged priority to provide enough security with less resource constraints for industrial applicability.

In the following background materials required to understand the proposed work are explored.

**1.1 Material and methods:**

In this section some background knowledge of steganography and steganalysis will be explored as a prerequisties of the proposed work.

**1.1.1 Steganography**

In the proposed work, the image is used as a steganography medium which can be classified as lossy and lossless. Lossy compression (for example JPEG design) accomplishes a significant level of compression and hence saves more space. However, doing so, the pieces might be adjusted generally and the inventiveness of the image might be affected. Whereas lossless compression reproduces the message precisely. We have implemented our algorithm with lossless compression image formats where the results have shown satisfactory levels of robustness. In our proposed scheme we have used spatial domain. In case of spatial, the focus is based on values of image pixels.

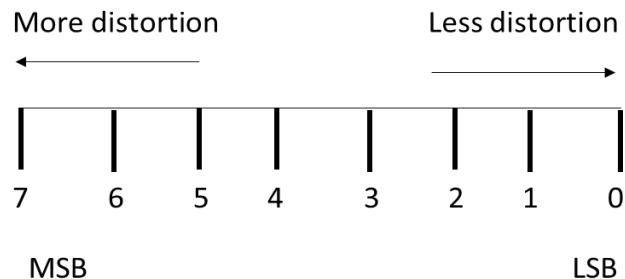

Fig 1. Embedding in 8-bit image

### 1.1.2 Steganalysis & Performance Measures

Steganalysis is the craftsmanship and science to recognize whether a given picture contains shrouded information. The steganalysis assumes a part in the determination of highlights or properties of the picture to test for concealed information and furthermore in planning of methods to distinguish or extricate the shrouded information. A steganalysis technique is considered effective if it can identify and extract the concealed information implanted. The proposed work mainly focuses on Targeted Steganalysis and Blind Steganalysis.

### 2. Proposed Work

Our proposed architecture is divided into two phases: Embedding phase and Extraction phase. In the proposed architecture, three main components are cover image, secret text to be embedded and the last one is direct key or questionnaire. For the questionnaire, the answer is inserted through the input at the beginning of the implemented program or a direct key can be used for encryption which is also overlaid on the stego-image. Here for the implementation the secret text and key are kept under ASCII format. For encryption, there are different cryptographic techniques already available. Among them, a very easily applicable caesar cipher or shift cipher is used in this paper. After inserting the cover image in the program, it checks for compression type. The algorithm is implemented to run in non-compression formatted images. After inserting the image, the program separates the channels R, G and B from the image as the system is implemented for RGB images. For each channel, the pixel value is ranging from 0-255. Now according to the algorithm, the last digit will be replaced, so the value of the last digit can be from 0-9 which also signifies the change will be maximum up to 4 bits. With only the last digit from each pixel, each channel is further serialized and visualized which is like the following waveform for better understanding.

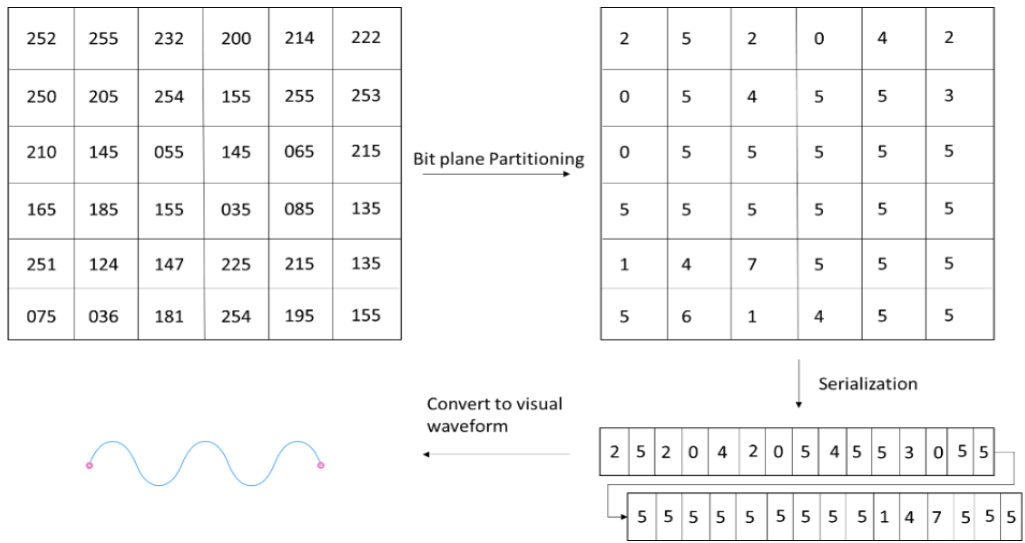

Fig.2 Image to waveform (Single channel)

In parallel, the secret text is also loaded and encoded into ASCII value list e.g., abcd -> 97-98-99-100. Now the answer or direct key, which was inserted, is used for encryption of the secret text. The serialized value string can be again visualized into waveform like Fig. 2. At this point, there are 4 serialized value strings (3 for RGB cover image and 1 for encrypted text). Now each serialized string from the cover image is fed to the differentiator with encrypted secret text's serialized value string. Where least standard deviation is calculated and that channel is selected for embedding to maximally obfuscate the secret text.

The embedding of the message is done (first layer security completed) and the rest is cryptographic key embedding which is the second layer. Most of the existing techniques embed the keys like a message so it can be extracted using the relatively same analyzing algorithm. Here we propose a special type of embedding where data is not embedded using its ASCII value, instead of that we have proposed a graphical character database for specific font size. Font size should be kept as small as possible to be non-visual under normal zoom level. The character database is formed using pixel coordinates and for overlaying first we choose the noisiest region of the image. The noisy region is chosen depending upon the number of different colors under a small block. After choosing the top, left position, using coordinates targeted pixels are given an offset of a specific constant value (e.g., we have given 10 for implementation). The overlaying should be done on a channel where secret text is not embedded; else secret text can be destroyed.

**2.1 Embedding Phase Proposed Architecture**

| EMBEDDING ALGORITHM |
|---|
| 1. Input cover image, secret text |
| 2. Compression method is checked for cover image |
| 3. If compression = lossless |
| 4.     Extract channels from image and serialize |
| 5.     Load secret text |
| 6.     Load character database |
| 7.     Input Questionnaire |
| 8.     Encrypt secret text using Caesar cipher and serialize |
| 9.     Check least standard deviation and select channel |
| 10.     If length == 1: /* length of rgb pixel value in form of string */ |
| 11.         New pixel value = new digit extracted from text digit |
| 12.         Make it 3 digit value (Example 3 -> 003) |
| 13.     Else if length == 2: /* length of rgb pixel value in form of string */ |
| 14.         Make it 3 digit value (Example 37 -> 037) |
| 15.         Value divided by 10 = Value without last digit |
| 16.     New pixel value = Value without last digit*10 + new digit extracted from text digit |
| 17.     Else if length == 3: /* length of rgb pixel value in form of string */ |
| 18.         Value divided by 10 = Value without last digit |
| 19.     New pixel value = Value without last digit*10 + new digit extracted from text digit |
| 20.     Text overlayed in done any channel except previously selected channel |
| 21.     Else /* compression = lossy */ |
| 22.         Not Implemented |

After embedding at sender's end, the extraction phase comes at the receiver end. After feeding the stego-image it is serialized. From each subsequent 3 groups of pixels last digit extracted and appended to form each letter of secret text as shown in fig 3

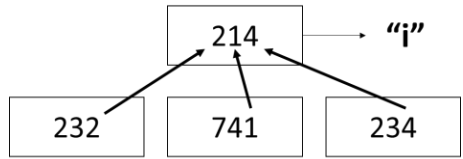

Fig.3 Secret text's character formation

But the characters were encrypted using Caesar encryption. So further the text is decrypted using the key manually entered in run-time.

## 2.2 Extraction Phase

**EXTRACTION ALGORITHM**

1. Take 3 subsequent image pixel values
2. Take An empty string
3. Loop through each pixel value
4.     Value modulo 10 = last digit extracted
5.     Append the digit to empty string
6.     Convert the string to integer which is in ASCII for the character
7. Enter the answer/key for Caesar decryption
8. Subtract the offset value (ASCII value of key) from each character to get the secret text

## 2.3 Proposed Architecture

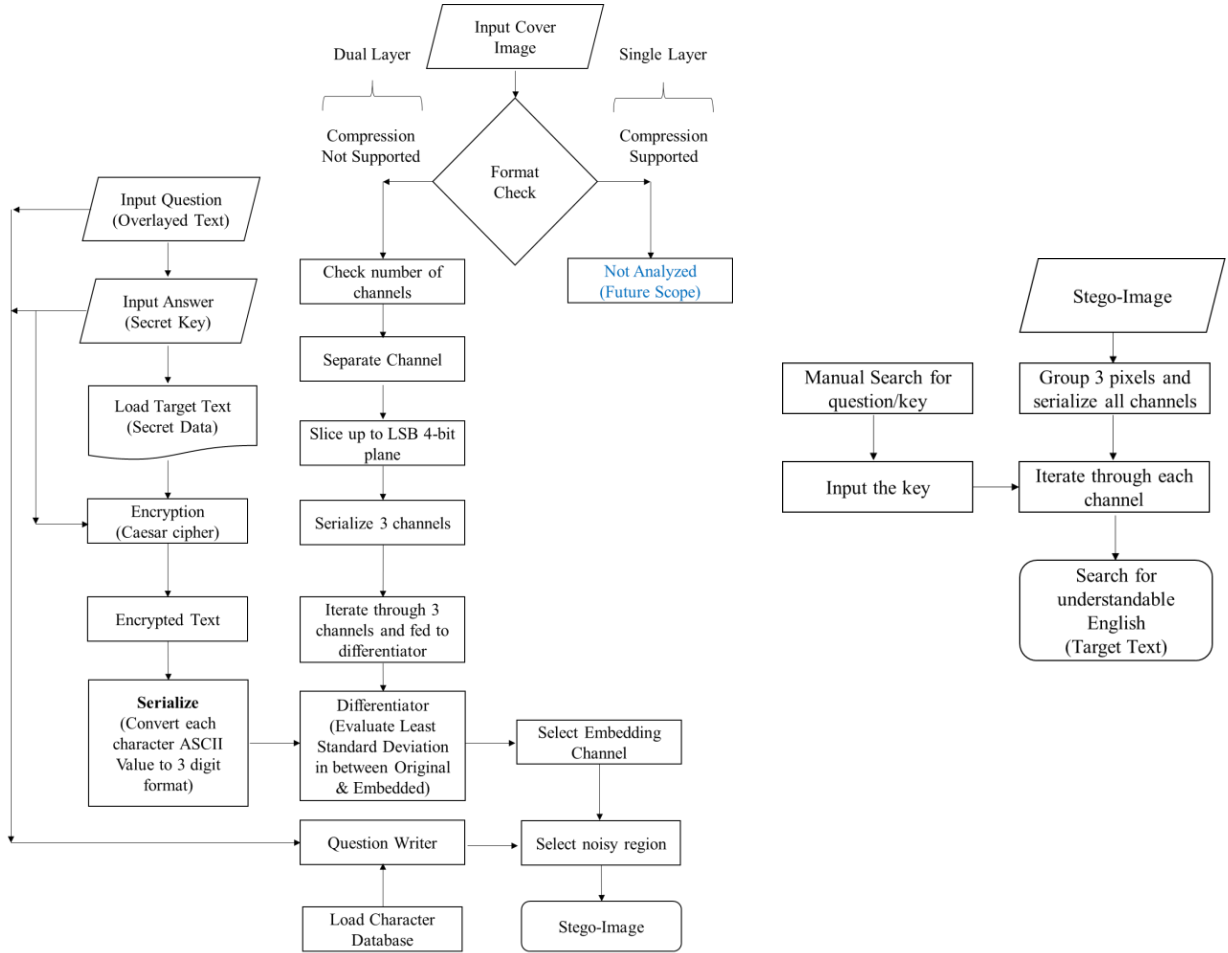

Fig 4(a). Embedding phase                                    Fig 4(b). Extraction phase

## 3. Implementation and Analysis:

In order to appraise the performance of our proposed scheme, we use difference performance & quality metrices like PSNR (Peak Signal to Noise Ratio), SSIM (Structural Similarity Index), standard images like Lena, Fruits, Splash etc. and standard datasets like CALTECH-BMP, USC-SIPI are considered. We presented all system parameters under the implemented scheme.

### 3.1 Tools and Computing system parameters

**Programming languages:** Python is used to build the software based on our proposed scheme and Java is used to run StegExpose [34] & Weka [33]. The tests were performed using Intel (R) Core (TM) i3-3210 with 3.20 GHz processor and 4.0 GB DDR3 RAM.

To execute from training to testing phase the following steps are executed.

**Step 1.** Steganography was done on all images from both data sets.
**Description**:
If the cover image name is cover0.bmp then the processed or stego-image can be saved as cover0-stego.bmp to discriminate both.

**Step 2.** Now first for doing the statistical steganalysis we have used StegExpose tool [34].
**Description**:
The tool can be downloaded from GitHub. After downloading there will be one folder namely "testFolder". We need to settle all the stego-images in that folder. Now to execute the StegExpose, first we need to open command line at same directory where StegExpose.jar is present. After that, we need to enter the following command.
**Command**: "java -jar StegExpose.jar test folder"

**Step 3.** Secondly, we have used Weka [33] for machine learning performance analysis.
**Description:**
After embedding we have listed all cover and stego images in a CSV file containing two columns namely Image name & Type. Image name is cover0.bmp, cover0-stego.bmp etc. and type is 0/1(Cover/Stego) respectively. After creating the list, we initiated the pretrained deep learning network Xception for feature extraction. For each image there are 2047 features are extracted. After feature extraction the generated CSV file was fed to Weka predefined classifiers under 10-fold cross-validation in the ratio of 67:33 (train:test).

In time of feeding a large amount of data Weka might be stopped as it will not be able to allocate the required amount of memory to process the data. So we have used the following command to start Weka from terminal and explicitly we have declared to allocate 1024 Megabytes of memory to java virtual machine for further processing.
**Command:** java -Xmx1024m -jar weka.jar

### 3.2 Dataset usage

In our implementation, we use two datasets for performance analysis namely CALTECH-BMP [30] and USC-SIPI Aerials, Miscellaneous Images [31]. USC-SIPI dataset was mainly used for quality analysis of steganography and CALTECH-BMP was used to check performance analysis of our proposed scheme against statistical analysis & machine learning classification through deep learning feature extraction.

### 3.3 Image size & Embedded data Relation

In our implementation, there were mainly two resolutions of 3 channel RGB images. For each resolution, we embedded a specific amount of data in one channel. In another channel, the graphical obfuscation was done and the rest channel was kept untouched. In the following Table 1. image size and embedded data size are mentioned.

Table 1. Size Factors of Image and Embedded data

| Image resolution (Width*Height*Channel) | Image size (pixels) | Embedded data size (bits) | Embedding Capacity (bits) |
|---|---|---|---|
| 256*256*3 | 196,608 | 152,000 | 1,048,576 |
| 512*512*3 | 786,432 | 648,000 | 4,194,304 |

### 3.4 Proposed scheme exceptional case

The proposed algorithm either obfuscates very well and in some cases the image is destroyed. This can be taken as a limitation of the proposed approach though the system can measure QoS for better performance in hiding.

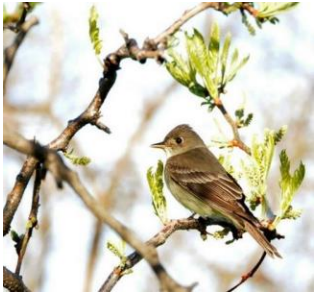 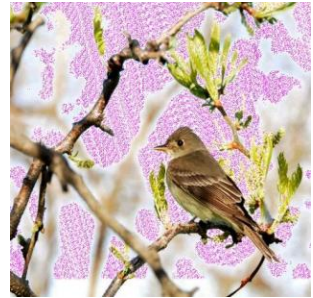

Fig 5(a). Original Cover      Fig 5(b). Stego-Image (Destroyed)

Table 2: First four records from total 1500 records (1500 images of CALTECH-BMP dataset)

| Stego-Image | Cover-Image | Similarity Index |
|---|---|---|
| 0_fmatted.bmp | C0001.bmp | 0.9965529646735172 |
| **1_fmatted.bmp** | **C0002.bmp** | **0.5974087739707975** |
| 2_fmatted.bmp | C0003.bmp | 0.9379443552261453 |
| 3_fmatted.bmp | C0004.bmp | 0.9816576154939605 |

From Table 2 it is proved that the second image is destroyed after embedding which was shown in the Fig.5(b).

### 4. Results and Discussion:

In this section, our proposed scheme performance factors are evaluated to check the effectiveness. Among the two datasets, USC-SIPI was mainly used to check effectiveness against statistical steganalysis, quality factor analysis and the other CALTECH-BMP dataset are used to check effectiveness against machine learning & statistical steganalysis.

### 4.1 Statistical analysis with USC-SIPI Dataset

In the following Table 3, we can notice one image from Aerial was detected under statistical steganalysis.

Table 3. Aerials tests under USC-SIPI dataset by StegExpose

| FileName | Crossed threshold? | Data Size | Primary-Sets | Chi-Square | Sample-Pairs | RS-analysis | Fusion-(mean) |
|---|---|---|---|---|---|---|---|
| 2.1.01._fmatted.bmp | FALSE | 38891 | 0.04776 | 0.00262 | 0.2667 | 0.1756 | 0.1483 |
| 2.1.02._fmatted.bmp | FALSE | 42712 | 0.04776 | 0.004061 | 0.2880 | 0.1966 | 0.1629 |
| 2.1.03._fmatted.bmp | FALSE | 46978 | 0.04776 | 0.001359 | 0.3333 | 0.2028 | 0.1791 |
| **2.1.04._fmatted.bmp** | **TRUE** | **53860** | **0.04776** | **0.002382** | **0.3686** | **0.2452** | **0.2054** |
| 2.1.05._fmatted.bmp | FALSE | 40202 | 0.04776 | 0.002492 | 0.2911 | 0.1663 | 0.1533 |
| 2.1.06._fmatted.bmp | FALSE | 31014 | 0.299953 | 0.004946 | 0.0127 | 0.1555 | 0.1183 |
| 2.1.07._fmatted.bmp | FALSE | 31257 | 0.0726 | 0.002183 | 0.2674 | 0.1346 | 0.1192 |
| 2.1.08._fmatted.bmp | FALSE | 39633 | 0.04776 | 0.004279 | 0.2872 | 0.1620 | 0.1511 |
| 2.1.09._fmatted.bmp | FALSE | 35431 | 0.04776 | 0.002176 | 0.2639 | 0.1392 | 0.1351 |
| 2.1.10._fmatted.bmp | FALSE | 30121 | 0.054389 | 0.002698 | 0.2612 | 0.1412 | 0.1148 |
| 2.1.11._fmatted.bmp | FALSE | 34402 | 0.04776 | 0.003985 | 0.2418 | 0.1478 | 0.1312 |
| 2.1.12._fmatted.bmp | FALSE | 40477 | 0.04776 | 0.002674 | 0.2899 | 0.1705 | 0.1543 |
| 2.2.01._fmatted.bmp | FALSE | 63903 | 0.015743 | 2.88E-04 | 0.1452 | 0.0824 | 0.0609 |
| 2.2.02._fmatted.bmp | FALSE | 44798 | 0.04776 | 7.61E-04 | 0.0905 | 0.0368 | 0.0427 |
| 2.2.03._fmatted.bmp | FALSE | 54288 | 0.04776 | 6.24E-04 | 0.1119 | 0.0427 | 0.0517 |
| 2.2.04._fmatted.bmp | FALSE | 45089 | 0.04776 | 6.51E-04 | 0.0836 | 0.0447 | 0.043 |
| 2.2.05._fmatted.bmp | FALSE | 48442 | 0.04776 | 9.12E-04 | 0.0955 | 0.0421 | 0.0461 |
| 2.2.06._fmatted.bmp | FALSE | 61995 | 0.014208 | 4.87E-04 | 0.1424 | 0.0792 | 0.0591 |
| 2.2.07._fmatted.bmp | FALSE | 62665 | 0.04776 | 7.78E-04 | 0.1207 | 0.0577 | 0.0597 |
| 2.2.08._fmatted.bmp | FALSE | 69578 | 0.04776 | 6.97E-04 | 0.1278 | 0.0705 | 0.0663 |
| 2.2.09._fmatted.bmp | FALSE | 41708 | 0.04776 | 3.97E-04 | 0.0842 | 0.0347 | 0.0397 |
| 2.2.10._fmatted.bmp | FALSE | 50626 | 0.04776 | 3.31E-04 | 0.0986 | 0.0458 | 0.0482 |
| 2.2.11._fmatted.bmp | FALSE | 47142 | 0.056992 | 5.83E-04 | 0.0909 | 0.0312 | 0.0449 |
| 2.2.12._fmatted.bmp | FALSE | 21177 | 0.021426 | 3.95E-04 | 0.0407 | 0.0182 | 0.0201 |
| 2.2.13._fmatted.bmp | FALSE | 52953 | 0.059059 | 5.15E-04 | 0.0999 | 0.0425 | 0.0504 |
| 2.2.14._fmatted.bmp | FALSE | 46047 | 0.047791 | 6.08E-04 | 0.0823 | 0.0448 | 0.0439 |
| 2.2.15._fmatted.bmp | FALSE | 53399 | 0.04776 | 6.69E-04 | 0.1066 | 0.0454 | 0.0509 |
| 2.2.16._fmatted.bmp | FALSE | 47790 | 0.033314 | 5.55E-04 | 0.0925 | 0.0559 | 0.0455 |
| 2.2.17._fmatted.bmp | FALSE | 43805 | 0.004362 | 0.001142 | 0.0979 | 0.0636 | 0.0417 |
| 2.2.18._fmatted.bmp | FALSE | 41475 | 0.02284 | 4.99E-04 | 0.0940 | 0.0407 | 0.0395 |
| 2.2.19._fmatted.bmp | FALSE | 52503 | 0.04776 | 7.51E-04 | 0.1043 | 0.0451 | 0.0500 |
| 2.2.20._fmatted.bmp | FALSE | 31143 | 0.00218 | 6.45E-04 | 0.0733 | 0.0426 | 0.0297 |
| 2.2.21._fmatted.bmp | FALSE | 55318 | 0.04776 | 7.23E-04 | 0.1013 | 0.0561 | 0.0527 |
| 2.2.22._fmatted.bmp | FALSE | 43966 | 0.03248 | 5.95E-04 | 0.0902 | 0.0443 | 0.0419 |
| 2.2.23._fmatted.bmp | FALSE | 37512 | 0.04776 | 6.00E-04 | 0.0743 | 0.0323 | 0.0357 |
| 2.2.24._fmatted.bmp | FALSE | 53343 | 0.03603 | 5.01E-04 | 0.1255 | 0.0414 | 0.0508 |

| wash-ir._fmatted.bmp | FALSE | 209249 | 0.04776 | 1.50E-04 | 0.0680 | 0.0558 | 0.0413 |

In the following figure we have represented PSNR of 37 images from Aerial USC-SIPI dataset after steganalysis.

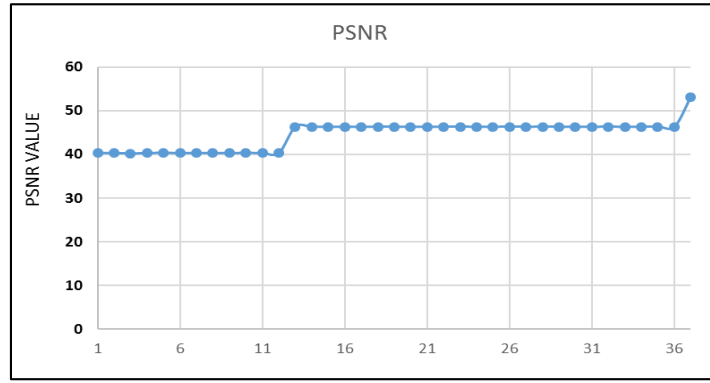

Fig 6. 37 image's PSNR under USC-SIPI dataset

Next, we have chosen some images from miscellaneous under USC-SIPI dataset. Those images were used to check other statistical properties like mean, median etc.

Table 4. MISC images test on statistical properties

| Sr. No. | Statistical Properties of Cover & Stego Images ||||
|---|---|---|---|---|
| 1 | 4.1.04 Female image properties ||||
| | **Attributes** | **Values** | Cover (TIFF) | Stego (BMP) |
| | Stego text | 19KB | | |
| | Above stego threshold? | FALSE | | |
| | Secret message size in bytes | 8726 | | |
| | Primary Sets | 0.080990432 | | |
| | Chi-square | null | | |
| | Sample Pairs | 0.153042101 | | |
| | RS analysis | 0.165308962 | | |
| | Fusion (mean) | 0.133113832 | | |
| | Capacity | 256*256*3 bytes | | |
| | FOBP | 0 | | |
| | | | **Attributes** / **Values** | **Attributes** / **Values** |
| | | | **Mean** 0.567 | **Mean** 0.461 |
| | | | **Median** 0.545 | **Median** 0.455 |
| | | | **Standard Deviation** 0.208 | **Standard Deviation** 0.213 |
| | | | **Pixels** 65536 | **Pixels** 196608 |
| 2 | 4.1.02 Female image properties ||||

| | Attributes | Values | Cover (TIFF) | Stego (BMP) |
|---|---|---|---|---|
| | Stego text | 19KB | 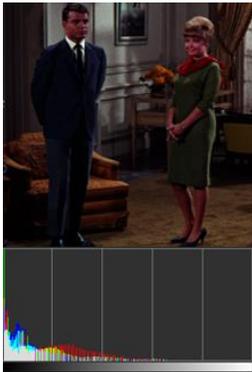 | 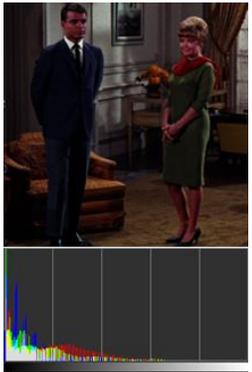 |
| | Above stego threshold? | FALSE | | |
| | Secret message size in bytes | 11645 | | |
| | Primary Sets | 0.080990432 | | |
| | Chi-square | null | | |
| | Sample Pairs | 0.224937686 | | |
| | RS analysis | 0.130343281 | | |
| | Fusion (mean) | 0.177640483 | | |
| | Capacity | 256*256*3 bytes | | |
| | FOBP | 0 | | |
| | | | Attributes | Values | Attributes | Values |
| | | | Mean | 0.173 | Mean | 0.129 |
| | | | Median | 0.137 | Median | 0.086 |
| | | | Standard Deviation | 0.140 | Standard Deviation | 0.129 |
| | | | Pixels | 65536 | Pixels | 196608 |
| 3 | | | Lena image properties | | | |
| | Attributes | Values | Cover (TIFF) | Stego (BMP) |
| | Stego text | 81KB | 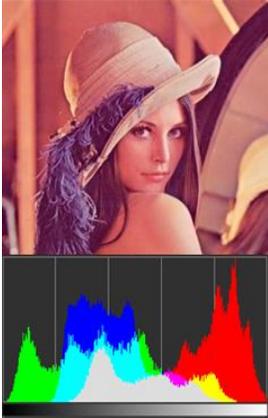 | 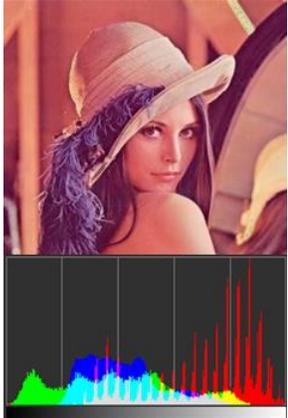 |
| | Above stego threshold? | FALSE | | |
| | Secret message size in bytes | 25885 | | |
| | Primary Sets | 0.018807427 | | |
| | Chi-square | 0.001913449 | | |
| | Sample Pairs | 0.239993506 | | |
| | RS analysis | 0.134234913 | | |
| | Fusion (mean) | 0.098737324 | | |
| | Capacity | 512*512*3bytes | | |
| | FOBP | 0 | | |
| | | | Attributes | Values | Attributes | Values |
| | | | Mean | 0.503 | Mean | 0.501 |
| | | | Median | 0.467 | Median | 0.467 |
| | | | Standard Deviation | 0.231 | Standard Deviation | 0.230 |
| | | | Pixels | 786432 | Pixels | 196608 |
| 4 | | | Fruits image properties | | | |

| | | Attributes | Values | Cover (TIFF) | | Stego (BMP) | |
|---|---|---|---|---|---|---|---|
| | | Stego text | 81KB | 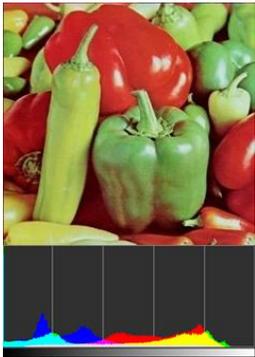 | | 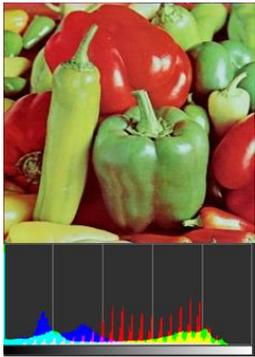 | |
| | | Above stego threshold? | FALSE | | | | |
| | | Secret message size in bytes | 44471 | | | | |
| | | Primary Sets | NaN | | | | |
| | | Chi-square | 0. 00144562 | | | | |
| | | Sample Pairs | 0. 136823965 | | | | |
| | | RS analysis | 0. 136062193 | | | | |
| | | Fusion (mean) | 0. 091443926 | | | | |
| | | Capacity | 512*512*3 | | | | |
| | | FOBP | 0 | | | | |
| | | | | Attributes | Values | Attributes | Values |
| | | | | Mean | 0. 434 | Mean | 0. 432 |
| | | | | Median | 0. 424 | Median | 0. 424 |
| | | | | Standard Deviation | 0. 260 | Standard Deviation | 0. 259 |
| | | | | Pixels | 786432 | Pixels | 786432 |
| 5 | | | | **Splash image properties** | | | |
| | | Attributes | Values | Cover (TIFF) | | Stego (BMP) | |
| | | Stego text | 81KB | 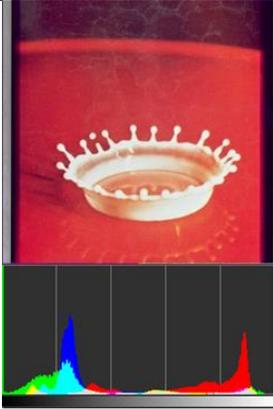 | | 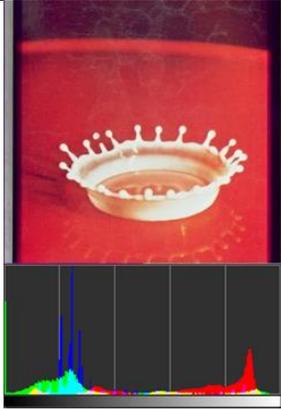 | |
| | | Above stego threshold? | FALSE | | | | |
| | | Secret message size in bytes | 23973 | | | | |
| | | Primary Sets | NaN | | | | |
| | | Chi-square | 0.00144562 | | | | |
| | | Sample Pairs | 0. 136823965 | | | | |
| | | RS analysis | 0. 136062193 | | | | |
| | | Fusion (mean) | 0. 091443926 | | | | |
| | | Capacity | 512*512*3 | | | | |
| | | FOBP | 0 | | | | |
| | | | | Attributes | Values | Attributes | Values |
| | | | | Mean | 0. 427 | Mean | 0. 425 |
| | | | | Median | 0. 271 | Median | 0. 271 |
| | | | | Standard Deviation | 0. 289 | Standard Deviation | 0. 289 |
| | | | | Pixels | 786432 | Pixels | 786432 |

## 4.2 Statistical & Machine Learning analysis with CALTECH-BMP Dataset

With CALTECH-BMP dataset there are different test scenarios which are derived in the following table 5.

Table 5. Implementation Scenarios

| As we have already discussed regarding the exceptional negative impact in section 3.4, we tested our algorithm in two cases or condition. At one condition after embedding image is exported even SSIM is less than 90% and in other case, images are exported only when SSIM is greater than equal to 90%. After embedding there were to test dataset containing all 1000 and remaining 832 images under worst and good condition respectively. $$f(x) = \begin{cases} Worst\ Case\ (1000\ images), & SSIM < 90\% \\ Good\ Case\ (832\ images)\ , & SSIM \geq 90\% \end{cases}$$ ||
|---|---|
| **Worst Case** ||
| **Case** | **Implemented Summary** |
| Statistical analysis | Total images for testing: 1000, 5.4% of images were correctly classified. |
| ML analysis with Deep learning | Total images for testing: cover (1000) + stego (1000) = 2000  Total features extracted: 2000*2047 = approx. 40 Lakhs 94 thousand, Classification result is presented at section 4.2.1 |
| **Good Case** ||
| **Case** | **Implemented Parameters** |
| Statistical analysis | Total images for testing: 832, 2.55% images were correctly classified. |
| ML analysis with Deep learning | Total images for testing: cover (832) + stego (832) = 1664  Total features extracted: 1664*2047 = approx. 34 Lakhs 6 thousand, Classification result is presented at section 4.2.2 |

### 4.2.1 Worst Condition (Destroyed images not filtered):

Table 6. Different ML classifier's result on multiple parameters in a worst condition

| Technique | TP Rate | FP Rate | Accuracy (Percentage) | Precision (Percentage) | ROC-Area (Percentage) |
|---|---|---|---|---|---|
| Naive Bayes | 0.65 | 0.34 | 65.2 | 66.5 | 71 |
| Random Forest | 0.71 | 0.28 | 71.1 | 71.1 | 78.1 |
| SMO | 0.735 | 0.265 | 73.5 | 73.5 | 73.5 |
| AdaBoostM1 | 0.66 | 0.33 | 66.3 | 66.3 | 73.1 |
| Bayes Net | 0.64 | 0.35 | 64.8 | 65.1 | 71.2 |
| Decision Table | 0.63 | 0.36 | 63.4 | 63.4 | 68 |
| **J48** | **0.74** | **0.25** | **74.3** | **74.3** | **81.2** |
| Logit-Boost | 0.66 | 0.34 | 66 | 66 | 74.1 |

In table 6. we can observe that the J48 classifier gave a maximum accuracy of 74.3% under the worst condition. Fig.7(a) show the ROC curves for each classifier for better representation.

### 4.2.2 Good Condition (Destroyed images filtered):

Table 7. Different ML classifier's prediction result on multiple parameters at good condition

| Technique | TP Rate | FP Rate | Accuracy (Percentage) | Precision (Percentage) | ROC-Area (Percentage) |
|---|---|---|---|---|---|
| Naïve Bayes | 0.35 | 0.64 | 35.75 | 31.6 | 29.8 |
| Random Forest | 0.08 | 0.92 | 7.99 | 8 | 1.7 |
| SMO | 0.219 | 0.781 | 21.87 | 21.9 | 21.9 |
| AdaBoostM1 | 0.46 | 0.53 | 46.39 | 46.3 | 44.2 |
| Bayes Net | 0.49 | 0.5 | 49.87 | 49.8 | 49.8 |
| **Decision Table** | **0.49** | **0.5** | **49.93** | **49.9** | **50** |
| J48 | 0.49 | 0.5 | 49.87 | 49.8 | 49.8 |
| Logit-Boost | 0.38 | 0.61 | 38.64 | 38.6 | 34.5 |
| SVM | 0.44 | 0.55 | 44.29 | 44.2 | 44.3 |

From the above result it is noticeable that in good condition maximum accuracy was only 49.93% using Decision table. Fig.7(b) depicts the ROC curves of different ML classifier's prediction results at good condition.

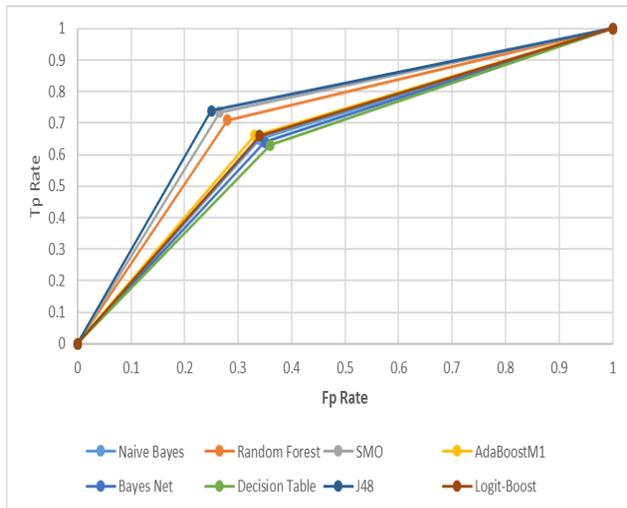
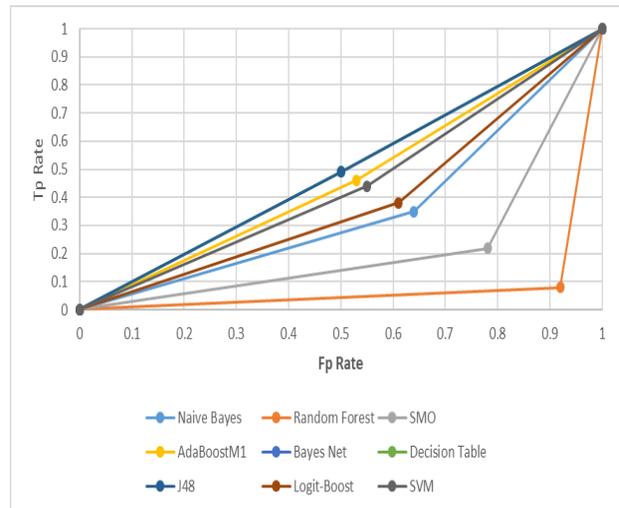

Fig 7(a). ROC curve of different ML classifier's prediction result at worst condition

Fig 7(b). ROC curve of different ML classifier's prediction result at good condition

### 4.3 Effectiveness in implementation

We demonstrated our scheme using different explanations and evaluated with more than one performance metrics. And in this last section of the result and discussion, we will discuss the effectiveness of our proposed scheme.
- Two layers of security: Obfuscation helps to meet the steganography objective and encryption layer makes the hidden data secured from an adversary.

- Here the secret text was taken from lorem-ipsum website [32]. The 81 KB of data was made by repeatedly copying the same portion of data from lorem-ipsum website. So, if the data is serialized then we will see a symmetrical waveform which also signifies that the secret text itself contains a pattern. This could be a very exploitation against hidden though the proposed scheme successfully could obfuscate it from get detected.
- From Fig.8 we can see the overlaid key is nearly visible after zooming and which cannot be extracted without human intervention. Standard OCRs might not be able to do key extraction as the font-size and visibility is very less. So, if the obfuscation layer is suspected then also with automation it is nearly infeasible to break the encryption layer.

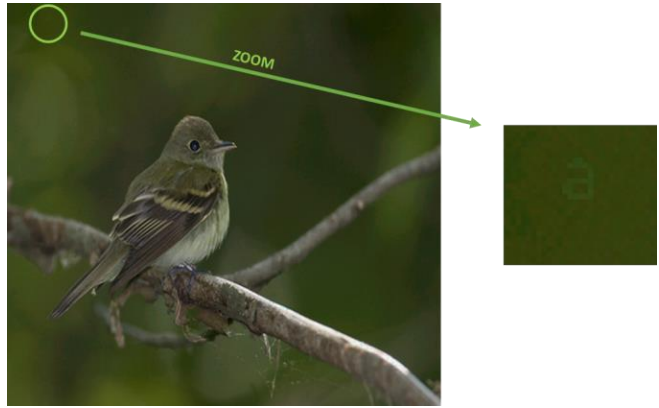

Fig 8. Visual embedding of answer or key

**5. Conclusion:**

We observe that deep learning in steganalysis is currently rising day by day. Taking this into consideration this paper proposed an efficient data hiding scheme. The analysis was done on different standard steganography datasets with standard machine learning classifiers, deep learning feature extractor, statistical measures. Explicitly the proposed work demonstrates quality analysis. From results it can be inferred that proposed steganography algorithm gives a significant performance and is capable of withstanding against statistical and ML steganalysis attacks. The proposed scheme is very light-weight so it can be used in different industrial security applications e.g., IoT communication under Industry 4.0 revolution. One more beneficial side of this research is, one layer of embedding was done in graphical space without ASCII value injection which is subjected to anti-automation attacks. As a future scope of the present work, instead of BMP images, GIF images as a dataset can be used to improve the capacity and security of the algorithm.

**Compliance with Ethical Standards:**

**Conflict of Interest:**
The authors declare that they have no conflict of interest.
**Ethical Approval:**
This article does not contain any studies with human participants or animals performed by any of the authors.
**Availability of data, material and code**

All required materials required to reproduce the results are available at the following google drive link.
https://drive.google.com/file/d/1MlRqMe01vxoAITG_3HUZTqlUWp6SMNVu/view?usp=sharing
This research did not receive any specific grant from funding agencies in the public, commercial, or not-for-profit sectors.

**Acknowledgements:**
We are grateful to the editor and other anonymous reviewers for their valuable response.

**References:**

[1] Zhang M, Zhang Y, Jiang Y, Shen J. Obfuscating eves algorithm and its application in fair electronic transactions in public cloud systems. IEEE System Journal. 2019;13(2):1478–1486. [Google Scholar]
[2] Cheng K, Wang L, Shen Y, Wang H, Wang Y, Jiang X, et al. Secure k-nn query on encrypted cloud data with multiple keys. IEEE Transactions on Big Data.2017.
[3] Wang H, Zhang Z, Taleb T. Special issue on security and privacy of IoT. World Wide Web. 2018;21(1):1–6. [Google Scholar]
[4] Ozcan, S., & Mustacoglu, A. F. (2018). Transfer Learning Effects on Image Steganalysis with Pre-Trained Deep Residual Neural Network Model. 2018 IEEE International Conference on Big Data (Big Data). doi:10.1109/bigdata.2018.8622437
[5] Khan, A., Siddiqa, A., Munib, S., and Malik, S. A. 2014. A recent survey of reversible watermarking techniques, Information Sciences 279 (2014), 251-272.
[6] Subhedar, M. S. and Mankar, V.H. 2014. Current status and key issues in image steganography: a survey, Computer Science Review 13 (2014), 95-113.
[7] Nissar, A., Mir, A. H. (2010). Classification of steganalysis techniques: a study, Digital Signal Processing 20 (2010), 1758-1770.
[8] Cho, S., Cha, B. H., Gawecki, M., and Kuo, C. C. 2013. Block-based image steganalysis: algorithm and performance evaluation, J. Vis. Commun. Image R. 24 (2013), 846-856.
[9] Karampidis, K., Kavallieratour, E., and Papadourakis, G. 2018. A review of image steganalysis techniques for digital forensics. Journal of Information Security and Applications 40 (2018), 217-235.
[10] Musumeci, F. et al. 2018. An overview on application of machine learning techniques in optical networks. Computer Science, Cornell University Library (Oct. 2018), 1-27. https://arxiv.org/abs/1803.07976
[11] https://www.infosecurity-magazine.com/magazine-features/the-dark-side-of-automation/
[12] Mielikainen J. LSB matching revisited. IEEE Signal Processing Letters. 2006;13(5):285–287. [Google Scholar]
[13] Zhang X, Wang S. Efficient Steganographic Embedding by Exploiting Modification Direction. IEEE Communications Letters. 2006;10(11):781–783. 10.1109/LCOMM.2006.060863
[14] Kim HJ, Kim C, Choi Y, Wang S, Zhang X. Improved modification direction methods. Computers and Mathematics with Applications. 2010;60(2):319–325. 10.1016/j.camwa.2010.01.006
[15] Chang C, Chou Y, Kieu T. An Information Hiding Scheme Using Sudoku. In: 2008 3rd International Conference on Innovative Computing Information and Control; 2008. p. 17–17.
[16] Hong W, Chen T. A Novel Data Embedding Method Using Adaptive Pixel Pair Matching. IEEE Transactions on Information Forensics and Security. 2012;7(1):176–184. 10.1109/TIFS.2011.2155062
[17] Chen J. A PVD-based data hiding method with histogram preserving using pixel pair matching. Signal Processing: Image Communication. 2014;29(3):375–384. 10.1016/j.image.2014.01.003
[18] Chang C, Liu Y, Nguyen T. A Novel Turtle Shell Based Scheme for Data Hiding. In: 2014 Tenth International Conference on Intelligent Information Hiding and Multimedia Signal Processing; 2014. p. 89–93.
[19] Liu L, Chang C, Wang A. Data hiding based on extended turtle shell matrix construction method. Multimedia Tools and Applications. 2017;76(10):12233–12250.
[20] Chang C, Liu Y. Fast turtle shell-based data embedding mechanisms with good visual quality. Journal of Real-Time Image Processing. 2018.
[21] Liu Y, Zhong Q, Chang L, Xia Z, He D, Cheng C. A secure data backup scheme using multi-factor authentication. IET Information Security. 2016;11(5):250–255.

[22] Martínez-González RF, Díaz-Mendez JA, Palacios-Luengas L, López-Hernández J, Vázquez-Medina R. A steganographic method using Bernoulli's chaotic maps. Computers and Electrical Engineering. 2016;54:435—449. 10.1016/j.compeleceng.2015.12.005

[23] Yadav GS, Ojha A. Chaotic system-based secure data hiding scheme with high embedding capacity. Computers and Electrical Engineering. 2018;69:447–460. 10.1016/j.compeleceng.2018.02.022

[24] Zaid Ibrahim Rasool Rasool, Dr. Mudhafar Al-Jarrah. "The Detection of Data Hiding in RGB Images Using Statistical Steganalysis". Thesis, CS, Middle East University

[25] Kamaldeep Joshi, Swati Gill, Rajkumar Yadav, "A New Method of Image Steganography Using 7th Bit of a Pixel as Indicator by Introducing the Successive Temporary Pixel in the GrayScale Image", Journal of Computer Networks and Communications, vol. 2018, Article ID 9475142, 10 pages, 2018. https://doi.org/10.1155/2018/9475142

[26] Rasras, Rashad & Alqadi, Ziad & Rasmi, Mutaz & Abu Sara, Mutaz. (2019). A Methodology Based on Steganography and Cryptography to Protect Highly Secure Messages. Engineering, Technology & Applied Science Research. 9. 3681-3684.

[27] J. Qin, Y. Luo, X. Xiang, Y. Tan and H. Huang, "Coverless Image Steganography: A Survey," in IEEE Access, vol. 7, pp. 171372-171394, 2019, doi: 10.1109/ACCESS.2019.2955452.

[28] Durafe A. & Patidar V. (2020). Development and analysis of IWT-SVD and DWT-SVD steganography using fractal cover. Journal of King Saud University - Computer and Information Sciences. doi:10.1016/j.jksuci.2020.10.008

[29] Qian Z., Huang C., Wang Z., Zhang X. (2019) Breaking CNN-Based Steganalysis. In: Pan JS., Lin JW., Sui B., Tseng SP. (eds) Genetic and Evolutionary Computing. ICGEC 2018. Advances in Intelligent Systems and Computing, vol 834. Springer, Singapore. https://doi.org/10.1007/978-981-13-5841-8_50

[30] Al-Jarrah, Mudhafar (2018), "RGB-BMP Steganalysis Dataset", Mendeley Data, V1, doi: 10.17632/sp4g8h7v8k.1

[31] The USC-SIPI Image Database. http://sipi.usc.edu/database/

[32] lipsum.com

[33] Weka, The workbench for machine learning. https://www.cs.waikato.ac.nz/ml/weka/

[34] StegExpose. https://github.com/b3dk7/StegExpose

[35] Alfredo Canziani, Adam Paszke, Eugenio Culurciello. "An Analysis of Deep Neural Network Models for Practical Applications". arXiv:1605.07678

[36] AryfandyFebryan, Tito WaluyoPurboyo and Randy ErfaSaputra. "Analysis of Steganography on TIFF Image using".Journal of Engineering and Applied Sciences 15 (2): 373-379, 2020, ISSN: 1816-949X

[37] Rashid, Aqsa & Rahim, Muhammad. (2016). Critical Analysis of Steganography " An Art of Hidden Writing ". International Journal of Security and Its Applications. 10. 259-282. 10.14257/ijsia.2016.10.3.24.

[38] N. Sharma and U. Batra, "A review on spatial domain technique based on image steganography," 2017 International Conference on Computing and Communication Technologies for Smart Nation (IC3TSN), Gurgaon, 2017, pp. 24-27, doi: 10.1109/IC3TSN.2017.8284444.

[39] S. G. Shelke and S. K. Jagtap, "Analysis of Spatial Domain Image Steganography Techniques," 2015 International Conference on Computing Communication Control and Automation, Pune, India, 2015, pp. 665-667, doi: 10.1109/ICCUBEA.2015.136.

[40] Mehdi Hussain, Ainuddin Wahid Abdul Wahab, Yamani Idna Bin Idris, Anthony T.S. Ho, Ki-Hyun Jung. Image steganography in spatial domain: A survey. Signal Processing: Image Communication, Volume 65, 2018, Pages 46-66, ISSN 0923-5965, https://doi.org/10.1016/j.image.2018.03.012.

[41] Ozcan, S., & Mustacoglu, A. F. (2018). Transfer Learning Effects on Image Steganalysis with Pre-Trained Deep Residual Neural Network Model. 2018 IEEE International Conference on Big Data (Big Data). doi:10.1109/bigdata.2018.8622437

[42] Jung, K.-H. (2019). A Study on Machine Learning for Steganalysis. Proceedings of the 3rd International Conference on Machine Learning and Soft Computing - ICMLSC 2019. doi:10.1145/3310986.3311000

[43] Lee, J. H., Shin, J., and Realff, M. J. 2018. Machine Learning: overview of the recent progresses and implications for the process systems engineering field. Computer and Chemical Engineering 114 (Oct. 2017), 111-121.

[44] Schmidhuber, J. 2015. Deep learning in neural networks: an overview. Neural Networks 61 (Oct. 2014), 85-117.

[45] Zhang M, Zhang Y, Su Y, Huang Q, Mu Y. Attribute-based hash proof system under learning-with-errors assumption in obfuscator-free and leakage-resilient environments. IEEE Systems Journal. 2015;11(2):1018–1026.

[46] Lee C, Chang C, Wang K. An improvement of EMD embedding method for large payloads by pixel segmentation strategy. Image and Vision Computing. 2008;26(12):1670—1676. 10.1016/j.imavis.2008.05.005.

[47] Pretrained Deep Neural Networks. https://in.mathworks.com/help/deeplearning/ug/pretrained-convolutional-neural-networks.html
[48] Francois Chollet. "Xception: Deep Learning with Depth Wise Separable Convolutions".

**Appendix:**

1. **Ceaser encryption key, input from user**

```
ques=input('Enter security key (one character) (a-zA-Z): ')
ques_val=ord(ques)
```

2. **Convert characters of string (target hidden data) to list of decimal values**

```
rdata=''
with open("target.txt") as f:
    rdata=f.read()

rdata_l=list(bytes(rdata.encode()))
rdata_len=len(rdata_l)
```

3. **RGB channels serialization**

```
for i in range(rows):
  for j in range(cols):
    blue.append(image[i,j][0])
    green.append(image[i,j][1])
    red.append(image[i,j][2])
```

4. **Data embedding in each channel**

```
if len(text)<=len(blue):
    temp_ind=0
    for word in range(len(text)):
        if len(str(blue[word]))>=2:
            temp_blue.append((10*int(blue[word]/10))+text[word])
            temp_ind=word
        else:
            temp_blue.append(text[word])
            temp_ind=word
    if temp_ind!=len(blue)-1:
        for a in range((temp_ind+1),len(blue)):
            temp_blue.append(blue[a])

    temp_ind=0
    for word in range(len(text)):
        if len(str(green[word]))>=2:
            temp_green.append((10*int(green[word]/10))+text[word])
            temp_ind=word
        else:
```

```
                temp_green.append(text[word])
                temp_ind=word
        if temp_ind!=len(green)-1:
            for a in range((temp_ind+1),len(green)):
                temp_green.append(green[a])

        temp_ind=0
        for word in range(len(text)):
            if len(str(red[word]))>=2:
                temp_red.append((10*int(red[word]/10))+text[word])
                temp_ind=word
            else:
                temp_red.append(text[word])
                temp_ind=word
        if temp_ind!=len(red)-1:
            for a in range((temp_ind+1),len(red)):
                temp_red.append(red[a])
```

5. **After data embedding standard deviation is checked for each channel**

```
result1 = np.subtract(blue,temp_blue)
result2 = np.subtract(green,temp_green)
result3 = np.subtract(red,temp_red)
```

6. **Graphical character embedding and stego-image export**

```
diff=0
for a in ques:
    val=ord(a)
    if val>=97 and val<=122:
        val=26+(val-97)
    elif val>=65 and val<=90:
        val=val-65

    minv=min([i[0] for i in characters[val]])
    maxv=max([i[0] for i in characters[val]])
    diff+=(maxv-minv)
    for b in characters[val]:
        if choose==1:
            image[b[1],b[0]-minv+diff+40][1]+=30
            image[b[1],b[0]-minv+diff+40][2]+=30
        elif choose==2:
            image[b[1],b[0]-minv+diff+40][0]+=30
            image[b[1],b[0]-minv+diff+40][2]+=30
        else:
            image[b[1],b[0]-minv+diff+40][1]+=30
            image[b[1],b[0]-minv+diff+40][0]+=30

cv2.imwrite('output/'+str(counter)+'_fmatted'+'.bmp',image)
```

```
    print('Exported in output directory')
```

## 7. Hidden data extraction from stego-image

```
def extract(filename):
 image = cv2.imread(filename)
 rows,cols,ch = image.shape
 data=''
 blue1=[]
 green1=[]
 red1=[]
 ques1=input('Enter security key: ')
 ques_val1=ord(ques1)
 print('key',ques_val1)

 for i in range(rows):
   for j in range(cols):
         blue1.append(image[i,j][0])
         green1.append(image[i,j][1])
         red1.append(image[i,j][2])
 j=1
 for i in zip(*[iter(red1)]*3):
         try:
                 # print(int(str((i[0]%10))+str((i[1]%10))+str((i[2]%10))))
                 data+=chr((int(str((i[0]%10))+str((i[1]%10))+str((i[2]%10)))-ques_val1))
         except:
                 pass
 print(data)
```

## 8. PSNR calculation

```
def psnr(path1,path2):
        original = cv2.imread(path1)
        contrast = cv2.imread(path2)
        d = cv2.PSNR(original, contrast)
        return d
```

## 9. SSIM calculation

```
def ssim_val(path1,path2):
        imageA = cv2.imread(path1)
        imageB = cv2.imread(path2)
        # convert the images to grayscale
        grayA = cv2.cvtColor(imageA, cv2.COLOR_BGR2GRAY)
        grayB = cv2.cvtColor(imageB, cv2.COLOR_BGR2GRAY)
        (score, diff) = compare_ssim(grayA, grayB, full=True)
        diff = (diff * 255).astype("uint8")
        return score
```